# In-situ EXAFS study on the thermal decomposition of $TiH_2$*

ZHOU Ying-Li (周英丽)[1, 2; 1)] ZHENG Li-Rong (郑黎荣)[1] CHU Sheng-Qi (储胜启)[1]
Wu Min (吴敏)[1, 2] AN Peng-Fei (安鹏飞)[1, 2] ZHANG Jing (张静)[1] HU Tian-Dou (胡天斗)[1; 2)].

[1] Institute of High Energy Physics, Chinese Academy of Sciences, Beijing 100049, China
[2] University of Chinese Academy of Sciences, Chinese Academy of Sciences, Beijing 100049, China

**Abstract** Thermal decomposition behaviors of $TiH_2$ powder under a flowing helium atmosphere and in a low vacuum condition have been studied by using in-situ EXAFS technique. By an EXAFS analysis containing the multiple scattering paths including H atoms, the changes of hydrogen stoichiometric ratio and the phase transformation sequence are obtained. The results demonstrate that the initial decomposition temperature is dependent on experimental conditions, which occurs, respectively, at about 300 and 400 °C in a low vacuum condition and under a flowing helium atmosphere. During the decomposition process of $TiH_2$ in a low vacuum condition, the sample experiences a phase change process: $\delta(TiH_2) \rightarrow \delta(TiH_x) \rightarrow \delta(TiH_x) + \beta(TiH_x) \rightarrow \delta(TiH_x) + \beta(TiH_x) + \alpha(Ti) \rightarrow \beta(TiH_x) + \alpha(Ti) \rightarrow \alpha(Ti) + \beta(Ti)$. This study offers a way to detect the structural information of hydrogen. A detailed discussion about the decomposition process of $TiH_2$ is given in this paper.

## 1 Introduction

Titanium hydride ($TiH_2$) has attracted wide attention not only because it is a potential hydrogen storage material, but also because it has many applications in other fields. For example, it was used as a source of pure $H_2$ [1] to synthesize other hydrides, or to form the ceramic glass for metal seals [2]. It was used as a bonding material to attach the diamond surface to metals [3]. Besides, it was also used in the synthesis of Ti-based alloys [4-6]. However, all these applications are dependent on the decomposition of $TiH_2$ or the release of hydrogen when a heat treatment is performed under certain atmosphere. Therefore, there have been many researches about the dehydrogenation of $TiH_2$ by using thermal analysis method combined with structural analysis. V. Bhosle et al. [7, 8] examined the dehydrogenation of $TiH_2$ by using DTA/TGA, XRD, and TEM. They observed that the dehydrogenation of $TiH_2$ experienced a two-step process such as $TiH_2 \rightarrow TiH_x \rightarrow \alpha$-Ti. But the structural details of the intermediate phase $TiH_x$ are still unclear. C. Jiménez et al. [9, 10] studied the decomposition of the as-received $TiH_2$ powder during an isothermal treatment with flowing or stationary argon atmosphere by using in-situ X-ray and neutron diffraction techniques. The phase transformation sequence of the as-received $TiH_2$ was obtained and expressed as: $\delta \rightarrow \delta + \alpha \rightarrow \delta + \alpha + \beta \rightarrow \alpha + \beta \rightarrow \alpha$ under a flowing argon atmosphere. While changing the atmospheric conditions from flowing to stationary argon, the sequence changed to: $\delta \rightarrow \delta + \alpha \rightarrow \delta + \alpha + \beta \rightarrow \alpha + \beta \rightarrow \beta$. Evidently, the phase transition sequence of $TiH_2$ depends on the atmospheric conditions. Specially, the phase

* Supported by National Natural Science Foundation of China (10875143)
1) E-mail: zhouyl@mail.ihep.ac.cn,

2) E-mail: hutd@mail.ihep.ac.cn,

transition sequence of $TiH_2$ under flowing helium or low vacuum is still unknown.

To better understand the decomposition process of titanium hydrides, the phase transformations of $TiH_2$ under flowing He or low vacuum are studied in this work by using in-situ extended X-ray absorption fine structure (EXAFS) technique. Due to the element selectivity and the local structure sensibility, EXAFS spectroscopy is a powerful technique compared with XRD, TEM, and neutron diffraction. More importantly, EXAFS offers possibly a way for the local structure detection of hydrogen atom [11].

## 2 Experiments

$TiH_2$ powder, 99% pure and 33 μm in diameter, was supplied by Alfa Aesar. Its X-ray diffraction (XRD) pattern was collected at room temperature with Cu Kα radiation. The structural refinement of the XRD pattern shows that Ti atoms form a face centered cubic (fcc) crystal structure (i.e. δ phase) with a lattice parameter of 4.453±0.002Å, in which hydrogen atoms occupy the tetrahedral interstitial sites. Based on the relationship between the lattice volume and the hydrogen content of $TiH_x$ reported by Yakel et al. [12] and J. Li et al [13], the estimated hydrogen content is approximately equal to 2.

A certain amount of $TiH_2$ was homogeneously mixed with BN powder (supplied by Tianjin Chemical Reagent Factory), then was pressed to a 10 mm diameter tablet for EXAFS measurement. The X-ray absorption edge height at Ti K-edge was optimized to 1, i.e. $\Delta\mu d \approx 1$, by controlling the physical thickness $d$ of the tablet. Ti K-edge EXAFS spectra were measured in transmission mode at Beamline 1W1B of Beijing Synchrotron Radiation Facility (BSRF). A Si (111) double-crystal monochromator was used to monochromatize the incident X-ray energy with an energy resolution ($\Delta E/E$) of about $2\times10^{-4}$. The storage ring was operated at 2.13 GeV with an electron current decreasing from 450 to 330 mA in the span time of 2 h.

Under a flowing He atmosphere, the sample was step by step heated to 100, 200, 300, 350, 400, 450, 500, 550, and 600°C for in-situ EXAFS measurements. Before each EXAFS measurement, the sample temperature was held half an hour at the corresponding temperature point for thermal equilibrium. The uncertainty of sample temperature was less than 1 K. The EXAFS spectra were collected with a normal scan mode which spent about 20 minutes for one spectrum. In a low vacuum condition, the Ti K-edge EXAFS spectra were also collected. In this case, $TiH_2$ sample was continuously heated to 800°C from the room temperature with a constant heating rate of 2°C/min. Because the dehydrogenation of $TiH_2$ is significantly faster in vacuum condition than under the flowing inert atmosphere [14], quick EXAFS (QEXAFS) [15] mode was adopted in a low vacuum condition. The collection time span for each EXAFS spectrum was shortened to 90s.

## 3 Results and discussion

### 3.1 Room temperature EXAFS data analysis

The EXAFS data were analyzed with the software packages IFEFFIT [16]. Based on the crystallographic data obtained from the room temperature XRD pattern of $TiH_2$, the effective backscattering amplitude and the phase shift were calculated by using the FEFF8 [17] program. The room temperature EXAFS spectrum of $TiH_2$ was used as reference to extract the effective amplitude reduction factor $S_0^2$ and the energy threshold shift $\Delta E_0$.

First, the room temperature EXAFS spectrum of $TiH_2$ was fitted with a single-scattering path of Ti-Ti. In the fitting, the Ti-Ti distance was fixed at $a/2^{0.5}$, $a$ is the lattice parameter (a=4.453Å). Fig. 1 (a) and 1 (b) show, respectively, the fitting curves with Ti–Ti single-scattering path in R-space and in k-space. It can be found that the fitting goodness is not reasonable either in R-space or in k-space, which means that including only Ti-Ti single-scattering path in the fittings is inappropriate. To improve the fitting quality, the triple-scattering paths Ti-Ti-H or Ti-H-Ti were tested to be included in the EXAFS fittings. Fig. 1 (c) and 1 (d) show the fitting curves including the triple-scattering paths in R-space and in k-space. Evidently, the fitting quality of EXAFS data is greatly improved. These results demonstrate that the multiple scattering paths must be contained in the EXAFS fitting of $TiH_2$. Therefore, the same fitting strategy is used for all EXAFS fittings.

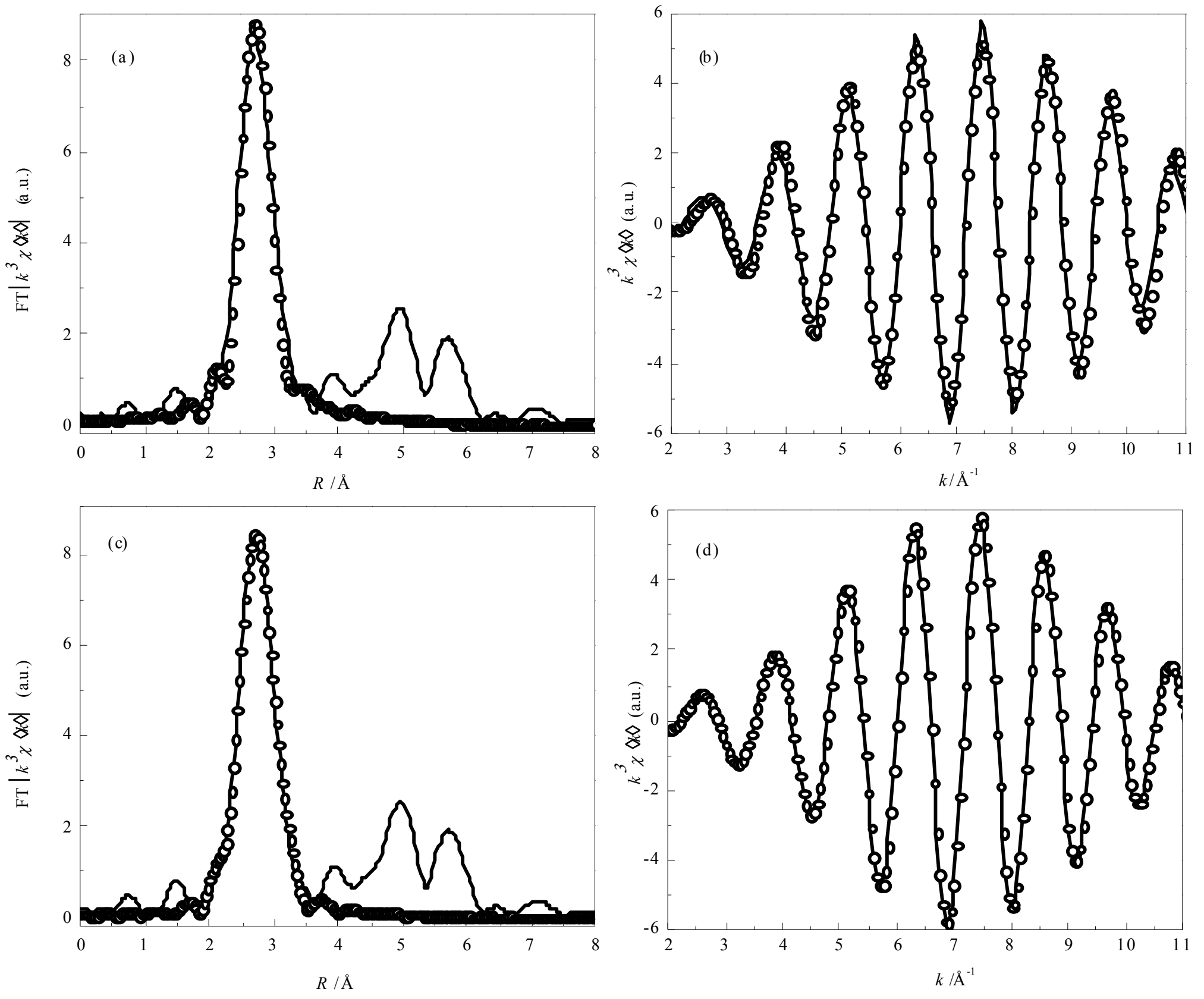

**Fig. 1**. Comparison of the Ti K-edge EXAFS fittings with and without triple-scattering paths for the room temperature $TiH_2$. The solid lines are the experimental values, and the hollow lines are the fitting curves. Fig.1(a) and 1(b) are the Fourier transform spectrum and its EXAFS oscillations fitted with only Ti-Ti path; Fig.1(c) and 1(d) are the Fourier transform spectrum and its EXAFS oscillations fitted with Ti-Ti path and Ti-Ti-H or Ti-H-Ti triple-scattering path together.

### 3.2 In-situ EXAFS data heated in flowing helium

The changes of Ti K-edge EXAFS oscillations with heating temperature are compared in Fig. 2 (a) for the $TiH_2$ sample under a flowing helium atmosphere. The corresponding Fourier transform spectra, without phase shift correction, are shown in Fig. 2 (b). The Fourier transform region is 2.5-11 $Å^{-1}$ in the k-space. From Fig. 2 (a), it can be seen that the EXAFS oscillation amplitude decreases obviously with the increasing temperature, but the oscillation features are almost unchangeable till 450°C. This result demonstrates that the local atomic structures around Ti in the $TiH_2$ sample are similar except an increase of the disorder when the sample temperature rises from room temperature up to 450°C. However, the EXAFS oscillation frequency has a little change when the sample temperature is up to 500°C, and presents obvious changes at 550 and 600°C compared with the lower temperature samples. All these imply that the local atomic structures have been changed above 500°C. From the Fourier transform spectra, it can be also found that the first main peak corresponds to the Ti-Ti distance of the first near neighbors. Obviously, the Ti-Ti distances shift to lower-R side and the magnitude of the main peak is greatly lowered above 500°C. Especially, a new coordination peak around 1.4 Å appears at 550 and 600°C. This coordination peak corresponds to the contribution of Ti-O distance, which indicates that the $TiH_2$ sample can be partially oxidized when the sample temperature is higher than 500°C.

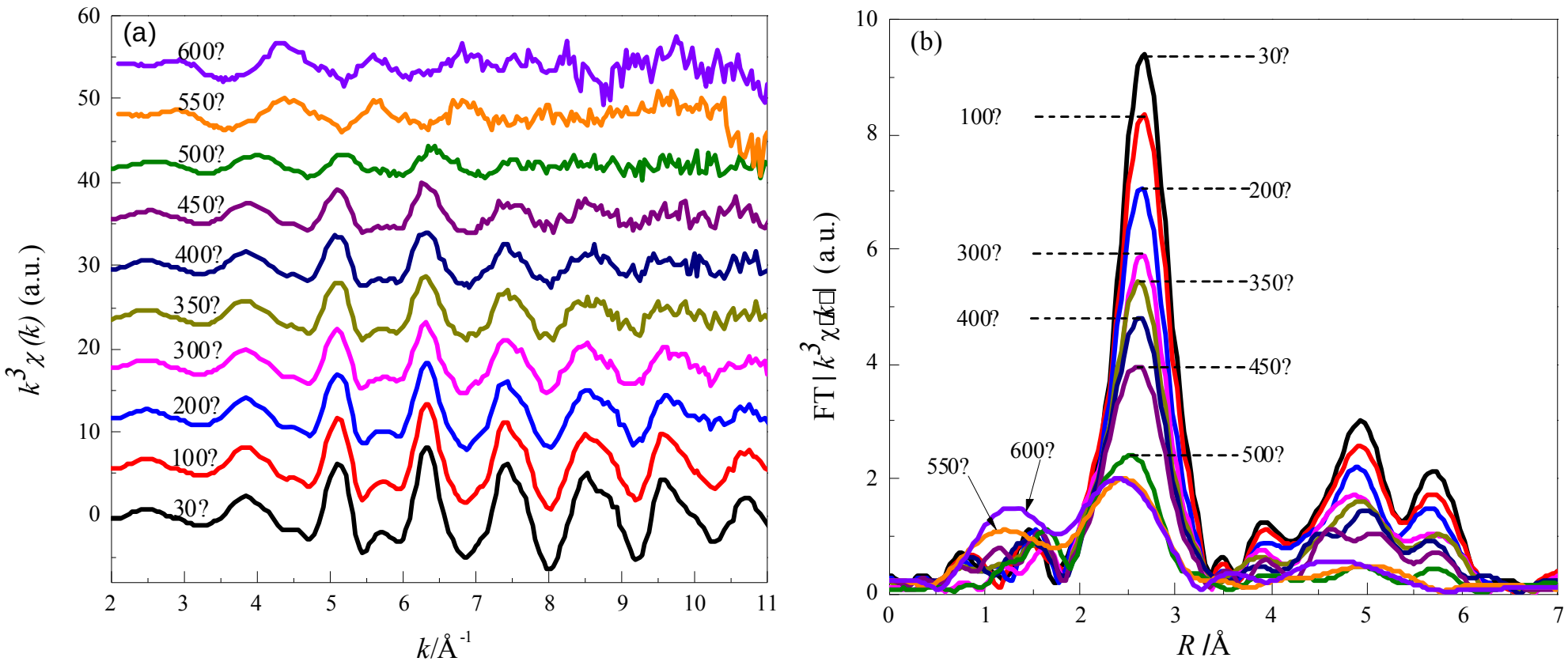

**Fig. 2.** The in-situ $k^3$-weighted EXAFS oscillations (a) and their Fourier transform spectra without phase shift correction in the k range of 2.5-11 Å$^{-1}$ (b) of the $TiH_2$ sample heated to different temperatures under a flowing helium atmosphere.

To get the structural change with heating temperature, the coordination peaks of Ti-Ti shell were isolated for the EXAFS fitting. As discussed above, the Ti-Ti single-scattering path together with the triple-scattering path were used in the EXAFS fittings. It is found that these EXAFS curves can be fitted very well with the sample temperature lower than 500°C when the coordination number and bondlength are fixed at the crystallographic data of the reference fcc structure, except a change of the Debye-Waller factor. However, for these EXAFS spectra with sample temperature higher than 450°C, they cannot be fitted well with the coordination number and bondlength fixed at the reference fcc structure. It reveals that the structure of $TiH_2$ has been changed with the release of hydrogen from the $TiH_2$ sample. If the structural constraint is released, the EXAFS fitting curves will be in excellent agreement with the experimental ones. On the other hand, the fitting parameters are also reasonable for these samples with temperature above 450°C. This result demonstrates that the decomposition of $TiH_2$ begins at 450°C under a flowing helium atmosphere.

The bondlength change of Ti-Ti with temperature is shown in Fig. 3 (a). The average thermal expansion coefficient is estimated to be $1.253\times10^{-5}$ K$^{-1}$ based on a linear fitting for the bondlengths in the temperature range of 30°C to 400°C. A drastic change of the bondlength in the temperature region from 450°C to 500°C implies a structural change of titanium hydride during the decomposition process. Although hydrogen is the lightest element and is almost transparent for X-rays, the EXAFS data analysis demonstrates that the multiple scattering path including hydrogen atoms has an important contribution to the EXAFS signal. In order to estimate the change of hydrogen content during the thermal decomposition of $TiH_2$, the hydrogen stoichiometric number (equals 2) in the room temperature titanium hydride is used as a standard, then the effective coordination number of the triple-scattering path is normalized to get the variation of stoichiometric numbers of hydrogen during the decomposition as shown in Fig. 3 (b). Obviously, a great release of H starts at 400°C. With a further increase of the heating temperature, the stoichiometric number of hydrogen decreases sharply.

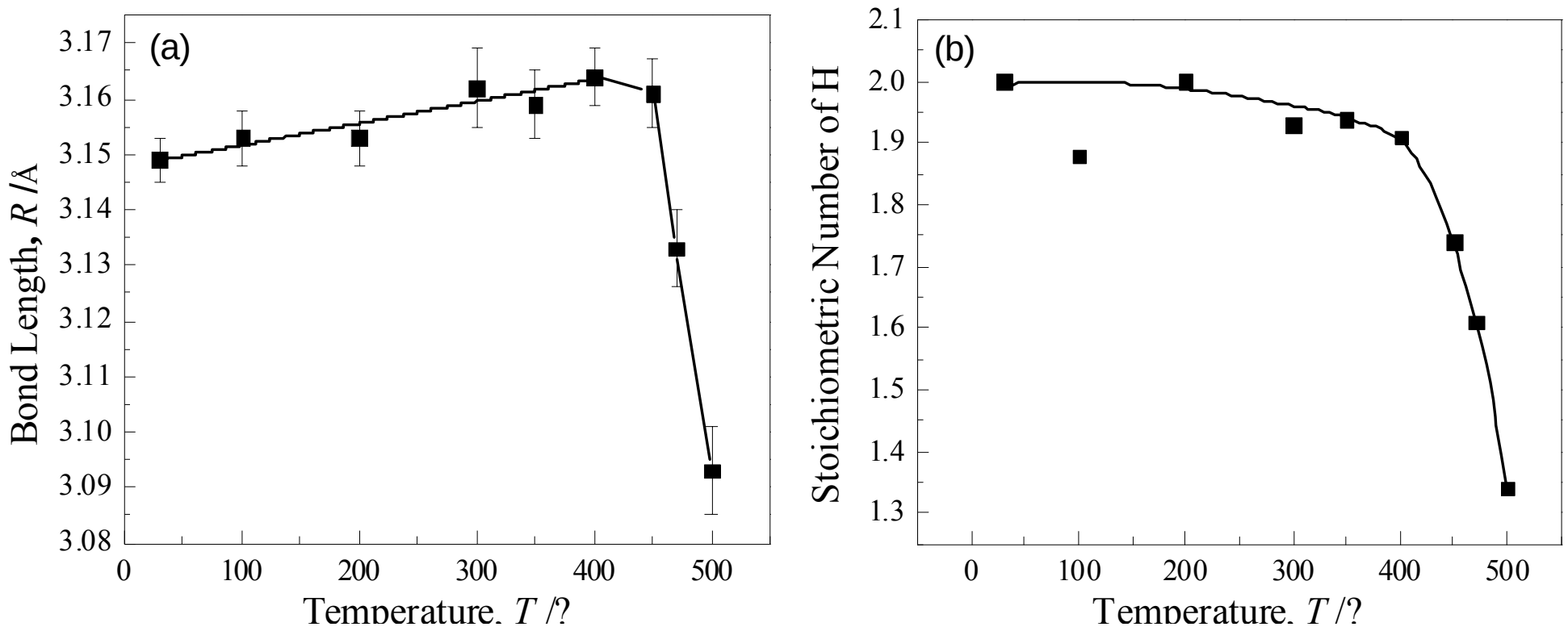

**Fig. 3**. Change of the EXAFS fitting parameters with heating temperature for the $TiH_2$ sample under a flowing helium atmosphere. Fig. 3 (a) is the bondlength change of Ti-Ti, Fig. 3(b) is the variation of the stoichiometric numbers of hydrogen in the titanium hydride.

### 3.3 In-situ EXAFS data heated in low vacuum

Quick EXAFS spectra were collected during the heating decomposition of $TiH_2$ in a low vacuum condition. A constant heating rate of 2°C/min was used. One EXAFS spectrum was collected per 3°C. Some representative QEXAFS oscillations with $k^3$-weight and their Fourier transform spectra without phase shift correction are shown in Fig. 4.

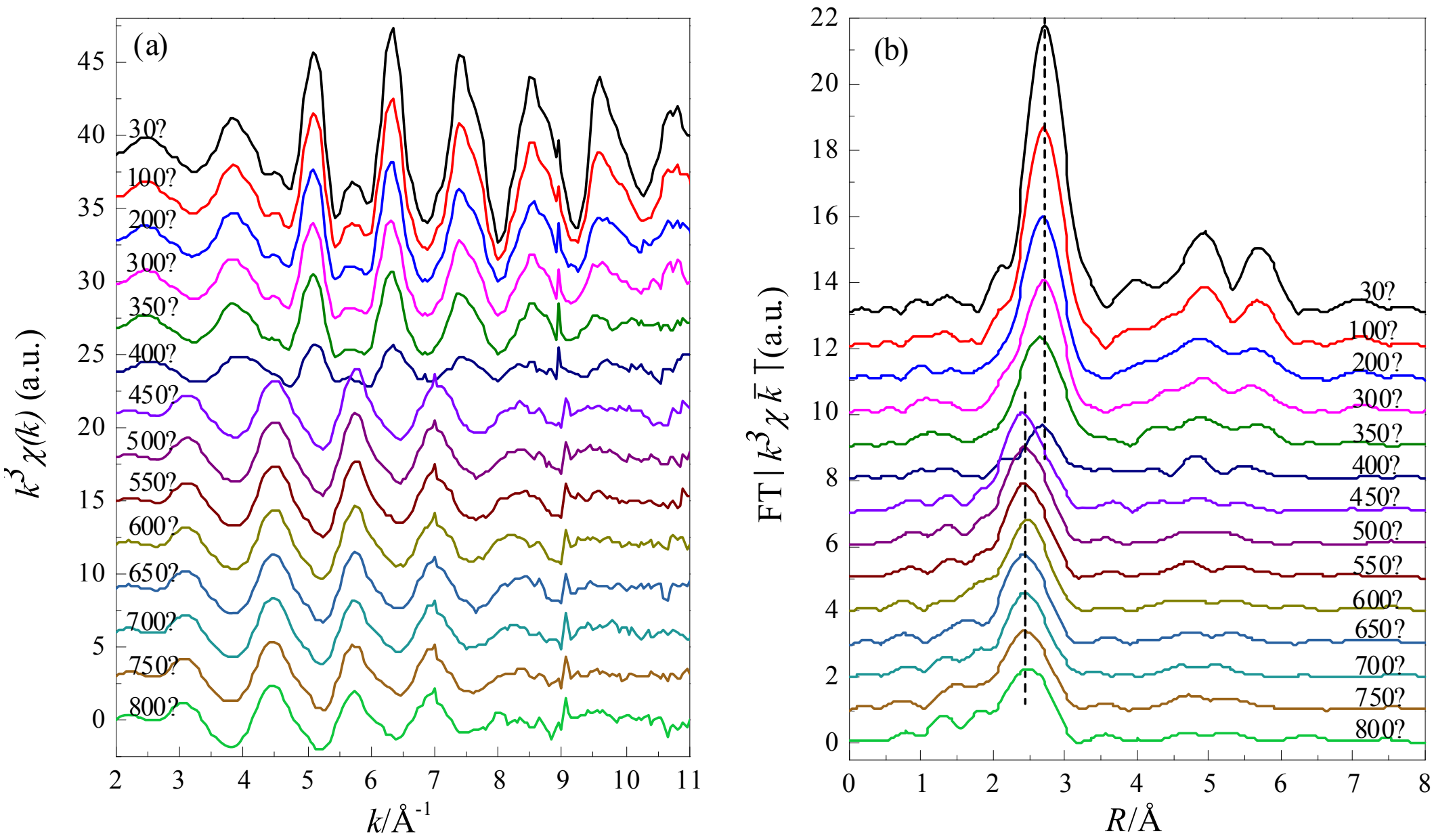

**Fig. 4.** Some representative QEXAFS oscillations with $k^3$-weight (a) and their Fourier transform spectra without phase shift correction (b) during the heating decomposition of $TiH_2$ sample in a low vacuum condition.

From Fig. 4, it can be easily observed that the QEXAFS oscillations and their Fourier transforms have an obvious alteration around 400°C, which implies that the structural transform happens between 350°C and 450°C. The main peaks in the Fourier transform spectra correspond to the nearest Ti-Ti bondlengths, i.e., the contribution of δ phase. From the experimental data, no obvious change can be observed from either the near edge absorption spectra or the Fourier transform spectra with temperature lower than 400°C. Therefore, the Ti-Ti shell was isolated from these QEXAFS spectra and was fitted by using the same fitting strategy as in the previous section. By the EXAFS fitting, it can be found that the $TiH_2$ sample can keep the fcc structure with sample temperature lower than 388°C in a low vacuum condition. The change of Ti-Ti bondlengths and the stoichiometric numbers of hydrogen in $TiH_x$ are shown in Fig. 5. Similarly, the Ti-Ti bondlength has a linear expansion up to 300°C with an average thermal expansion coefficient of $1.257\times10^{-5}\,K^{-1}$. This thermal expansion coefficient is almost the same as in the sample heated under a flowing helium atmosphere. This result demonstrates that the fcc structures are the same, without the dehydrogenation, in the linear expansion region for the sample under two different atmospheric conditions. The difference is that the fcc structure can keep up to 400°C under a flowing helium atmosphere, but it can be kept only up to 300°C in a low vacuum condition. In the linear expansion region, the stoichiometric numbers of hydrogen have no change and equal 2. When the sample temperature is higher than 300°C, due to the release of hydrogen from the sample, the Ti-Ti bondlength begins to contract. However, the samples still keep a δ-$TiH_x$ phase in the range of $1.3 < x < 2$. Beginning from 391°C, the fitting parameters cannot be attributed to an fcc structure. In other words, the samples are no longer a single δ-$TiH_x$ phase. Probably, certain new phase has formed when the sample temperature is higher than 391°C in a low vacuum condition.

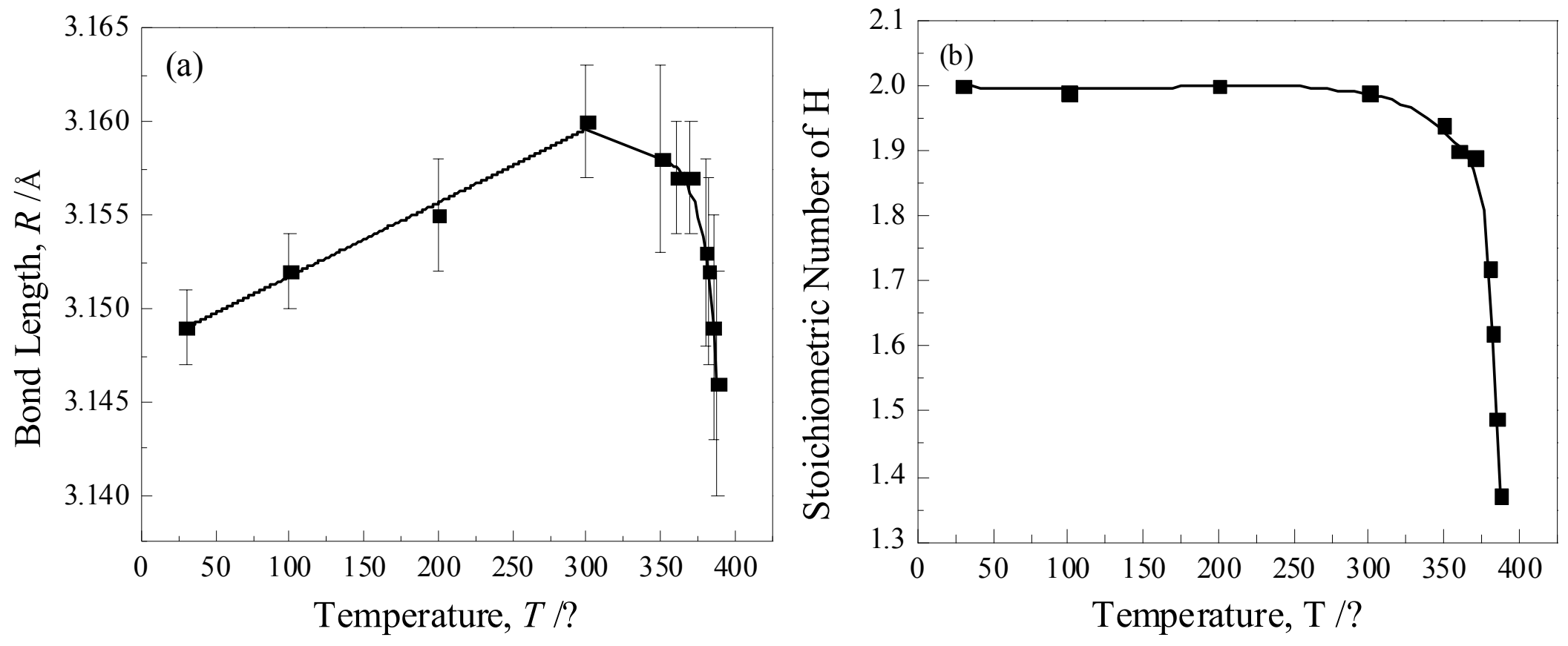

**Fig. 5.** Change of the EXAFS fitting parameters with heating temperature for the $TiH_2$ sample in a low vacuum condition. Fig. 5(a) is the bondlength change of Ti-Ti, Fig. 5(b) is the variation of the stoichiometric numbers of hydrogen in the titanium hydride.

Based on the Ti-H phase diagram [18] and literature [19-21], the new phase may belong to one of the following three phases: 1) α-Ti phase, which has a hexagonal close-packed (hcp) lattice, a=2.95 Å, c=4.69 Å; 2) β-$TiH_x$ phase, which is a body centered cubic (bcc) lattice, a=3.305 Å; 3) γ-$TiH_x$ phase, which is a face centered tetragonal (fct) lattice, a=4.21 Å, c=4.6 Å. The γ-$TiH_x$ phase is a metastable phase and exists only in a narrow range of hydrogen concentration ($0.79 \leqslant x \leqslant 1.0$). It was reported [22-24] that γ-$TiH_x$ phase was decomposed as the temperature was above 473K. Therefore, the possibility of the γ-$TiH_x$ appearance above 391°C is very small. If the new phase is an α-Ti phase, then there will be a shoulder appearance at the K-edge absorption spectra. Fig. 6 (a) shows the K-edge absorption spectra, but such a shoulder cannot be found at temperature around 391°C. Therefore, the possibility that the new phase is the α-Ti phase is also excluded. Consequently, the new phase appearing at 391°C has to be ascribed to the β-$TiH_x$ phase. In fact, the sample at 391°C is a mixture of δ-$TiH_x$ and β-$TiH_x$.

The quick X-ray absorption spectra and the Fourier transforms of QEXAFS spectra in a temperature range from 350°C to 450°C are shown in Fig. 6. With the temperature increasing to 407°C, the content of β-$TiH_x$ increases in the sample, which can be confirmed by the splitting of the Ti-Ti coordination peak or the appearance of another lower-R peak (R～2.2Å) as shown in Fig. 6 (d). Beginning from 411°C, the quick X-ray absorption spectra have a more and more significantly small shoulder on the absorption edge as shown in Fig. 6 (b), which indicates the appearance of an α-Ti phase. Therefore, the sample consists of δ-$TiH_x$, β-$TiH_x$ and α-Ti above 411°C.

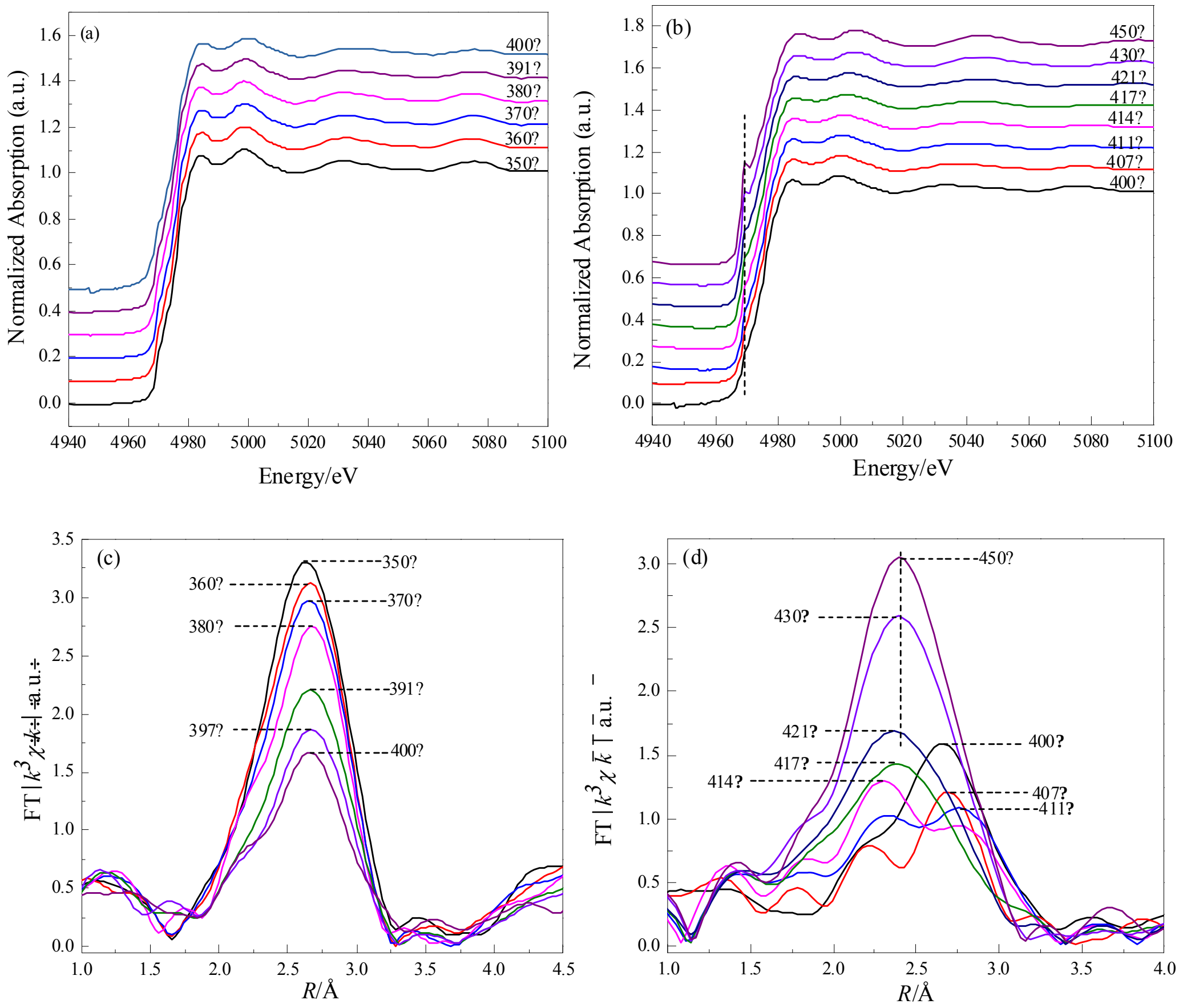

**Fig. 6.** The quick X-ray absorption spectra obtained by QEXAFS scan (a & b) and Fourier transform spectra of the corresponding QEXAFS signals (c & d) in a temperature range from 350 to 450°C.

The coordination peak at R～2.65Å is the contribution of Ti-Ti bonds in δ-phase as shown in Fig. 6 (c). When the temperature reaches up to 417°C, this peak disappears, which reveals that the δ-$TiH_x$ phase has been nearly decomposed. At this moment, the sample is composed of α-Ti and β-$TiH_x$. As the temperature further rises, β-$TiH_x$ is continuously dehydrogenated and transformed to α-Ti phase, which can be identified by the magnitude increase of the coordination peak in the Fourier transform spectra as shown in Fig. 6 (d). Accompanying the magnitude increase of the Ti-Ti coordination peak, its peak position shifts gradually from ~2.2 to ~2.4 Å at 430°C. This illustrates that the Ti-Ti bond length is longer in α-Ti than in β-$TiH_x$.

Above 430°C, the Fourier transform spectra are quite similar, and the peak positions of Ti-Ti coordination are consistent without any shift as shown in Fig. 4 (b) and Fig. 6 (d). It implies that there is no occurrence of phase change above 430°C. Perhaps, β-$TiH_x$ phase has completely transformed into α-Ti phase by dehydrogenation, or the survival β-$TiH_x$ can retain the bcc structure unchangeable in the subsequent heating process. For clarifying the high temperature structure of titanium hydride above 430°C, another experiment was performed under the same experimental conditions. The sample was heated to 430°C, and then cooled to room temperature. A room temperature EXAFS spectrum of the annealing sample was collected and compared with that of a standard Ti foil. Fig. 7 compares the Fourier transform spectra of the Ti foil, the annealing sample, and the high temperature samples. It can be found that the peak positions of Ti-Ti bond are almost the same between the annealing sample and the Ti foil. The result demonstrates that the annealing sample at 430°C has completely transformed into α-Ti. Actually, if β-$TiH_x$ was still kept in the sample after 430°C annealing, it would be transformed into δ-phase as the sample was cooled to room temperature. There is no evidence of δ-phase appearance in the annealing sample, which confirms that β-$TiH_x$ does not exist after a thermal treatment at 430°C in a vacuum condition.

From Fig. 7, it can be also found that the high temperature (≥ 450°C) samples have a shorter Ti-Ti bondlength compared with the Ti foil. This reflects that these high temperature samples are not the pure α-Ti phase. Usually, the metallic bondlength increases with the temperature increasing. However, the room temperature Ti-Ti bondlength in the annealing sample is longer than the high temperature Ti-Ti bondlength before annealing. This "abnormal" phenomenon can be ascribed to the phase transition from β-Ti to α-Ti as in metallic Ti. We believe that β-Ti exists in the high temperature samples. The lattice constant of β-Ti at 908°C was reported to be 3.3044 Å [25], which was used as a reference in the QEXAFS fitting with temperature higher than or equal to 450°C. The fitting result of the Ti-Ti bondlength was around 2.88 Å. It is between the Ti-Ti bondlengths of β-Ti (2.864 Å) and α-Ti (2.92 Å). Therefore, the samples with temperature above 450°C can be considered as a coexistence of α-Ti and β-Ti phases.

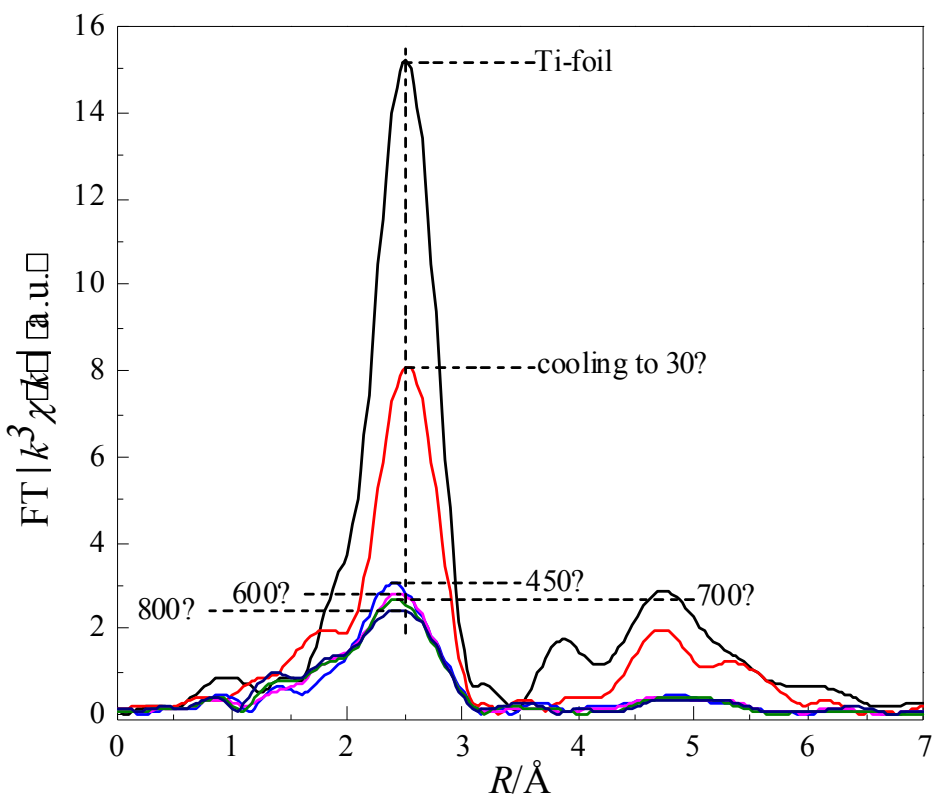

**Fig. 7.** Comparison of the Ti K-edge Fourier transform spectra among the Ti foil, the annealing sample at 430°C, and the high temperature samples with temperature above 450°C.

## 4 Conclusions

Thermal decomposition processes of $TiH_2$ under a flowing helium atmosphere and in a low vacuum condition were, respectively, monitored by in-situ EXAFS technique. The conclusions can be summarized as follows. The average thermal expansion coefficient of $TiH_2$ is $1.255\times10^{-5}$ $K^{-1}$ before the thermal decomposition. The initial decomposition temperature is above 400°C under a flowing helium atmosphere and is about 300°C in a low vacuum condition. The phase transformation sequence during the thermal decomposition process in a low vacuum condition is as follows:

$$\delta(TiH_2)\xrightarrow{300^\circ C}\delta(TiHx)\xrightarrow{391^\circ C}\delta(TiHx)+\beta(TiHx)\xrightarrow{411^\circ C}\delta(TiHx)+\beta(TiHx)+\alpha(Ti)$$
$$\xrightarrow{417^\circ C}\beta(TiHx)+\alpha(Ti)\xrightarrow{430^\circ C}\beta(Ti)+\alpha(Ti)$$

The contribution to the EXAFS spectra of hydrogen atoms can be indirectly extracted by considering the multiple scattering paths including H, which offers a way for the local structure detection of hydrogen atom. The stoichiometric ratio of hydrogen atoms in the titanium hydride can be quantitatively analyzed through the EXAFS fitting. The variations of stoichiometric ratio and phase composition during the decomposition process of $TiH_2$ can be used to predict the dehydrogenation rate.